\def\bold#1{\setbox0=\hbox{$#1$}%
     \kern-.025em\copy0\kern-\wd0
     \kern.05em\copy0\kern-\wd0
     \kern-.025em\raise.0433em\box0 }
\def\slash#1{\setbox0=\hbox{$#1$}#1\hskip-\wd0\dimen0=5pt\advance
       \dimen0 by-\ht0\advance\dimen0 by\dp0\lower0.5\dimen0\hbox
         to\wd0{\hss\sl/\/\hss}}
\documentstyle[12pt]{article}

\newlength{\dinwidth}
\newlength{\dinmargin}
\newcommand{\resection}[1]{\setcounter{equation}{0}\section{#1}}

\begin{document}

\def\lq{\left [}
\def\rq{\right ]}
\def\LL{{\cal L}}
\def\VV{{\cal V}}
\def\AA{{\cal A}}

\newcommand{\be}{\begin{equation}}
\newcommand{\ee}{\end{equation}}
\newcommand{\bea}{\begin{eqnarray}}
\newcommand{\eea}{\end{eqnarray}}
\newcommand{\nn}{\nonumber}
\newcommand{\dd}{\displaystyle}

\thispagestyle{empty}
\vspace*{4cm}
\begin{center}
  \begin{Large}
  \begin{bf}
EFFECTIVE LAGRANGIAN FOR HEAVY AND LIGHT MESONS: SEMILEPTONIC
DECAYS$^*$\\
  \end{bf}
  \end{Large}
  \vspace{5mm}
  \begin{large}
R. Casalbuoni\\
  \end{large}
Dipartimento di Fisica, Univ. di Firenze\\
I.N.F.N., Sezione di Firenze\\
  \vspace{5mm}
  \begin{large}
A. Deandrea, N. Di Bartolomeo and R. Gatto\\
  \end{large}
D\'epartement de Physique Th\'eorique, Univ. de Gen\`eve\\
  \vspace{5mm}
  \begin{large}
F. Feruglio\\
  \end{large}
Dipartimento di Fisica, Univ.
di Padova\\
I.N.F.N., Sezione di Padova\\
  \vspace{5mm}
  \begin{large}
G. Nardulli\\
  \end{large}
Dipartimento di Fisica, Univ.
di Bari\\
I.N.F.N., Sezione di Bari\\
  \vspace{5mm}
\end{center}
  \vspace{2cm}
\begin{center}
UGVA-DPT 1992/11-790\\\
BARI-TH/92-125\\\
November 1992
\end{center}
\vspace{1cm}
\noindent
$^*$ Partially supported by the Swiss National Foundation
\newpage
\thispagestyle{empty}
\begin{quotation}
\vspace*{5cm}
\begin{center}
  \begin{Large}
  \begin{bf}
  ABSTRACT
  \end{bf}
  \end{Large}
\end{center}
  \vspace{5mm}
\noindent
We introduce an effective lagrangian including negative and positive parity
heavy mesons containing a heavy quark, light pseudoscalars, and light vector
resonances, with their allowed interactions, using heavy quark spin-flavour
symmetry, chiral symmetry, and the hidden symmetry approach for light vector
resonances. On the basis of such a lagrangian, by considering the allowed weak
currents and by including the contributions from the nearest unitarity poles we
calculate the form factors for semileptonic decays of $B$ and $D$ mesons into
light pseudoscalars and light vector resonances. The available data, together
with some additional assumptions, allow for a set of predictions in the
different semileptonic channels, which can be compared with those following
{}from different approaches. A discussion of non-dominant terms in our
approach,
which attempts at including a rather complete dynamics,
will however have to wait till more abundant data become available.
\end{quotation}

\newpage
\setcounter{page}{1}
\resection{Introduction}

In this letter we shall present an analysis of semileptonic heavy
meson decays into light hadrons
\bea
P & \to &  \Pi \ell {\bar \nu}_l\\
P & \to &  \Pi^* \ell {\bar\nu}_l
\eea
($P$= heavy pseudoscalar
 meson, $\Pi$ and $\Pi^*$ = pseudoscalar and vector light mesons),
based on the use of heavy quark spin flavour symmetry \cite {Isgur},
chiral symmetry, and the hidden symmetry approach for light vector resonances.
Specifically, our framework will make use of: (i) the heavy-light chiral
lagrangian proposed in refs. \cite {Wise} \cite {Burd} \cite {Cho}
\cite {Yan}, which describes the
interaction of the pseudoscalar mesons belonging to the low-lying $SU(3)$
octet and the negative parity $J^P=0^-, 1^-$ heavy $Q{\bar q}$ mesons;
(ii) the introduction through the hidden gauge symmetry approach of the
vector meson resonances belonging to the low-lying $SU(3)$ octet within
the heavy-light chiral lagrangian \cite{light}; (iii) the inclusion
of low lying positive parity $Q{\bar q}$ heavy meson states within
the formalism.

We shall first summarize the well known description of the
interactions of heavy mesons and light pseudoscalars in terms of an effective
chiral lagrangian. In such a lagrangian we shall add a term describing the
octet vector meson resonances and their interactions with the heavy mesons and
the light pseudoscalars. We shall then introduce the effective lagrangian
containing the low-lying positive parity heavy meson states and their
interactions with the light pseudoscalars, with the negative parity heavy meson
states, and the couplings of the light vector resonances of the octet to both
positive and negative parity heavy meson states.

Unavoidably such an effective
description, will require the introduction of a set of coupling constants. The
study of the semileptonic decays (1.1) and (1.2) will be shown to yield some
information on such constants. To this end one has to use all the symmetry
constraints to characterize the form of the effective weak interaction of a
heavy negative or positive parity meson with the light pseudoscalars and of a
negative parity heavy meson with the light vector resonances. For a first
numerical analysis of the leptonic decays we shall be forced to neglect higher
derivative terms, which is justified only in limited portions of phase space.
After having introduced such a formal setting we shall analyze the semileptonic
decays (1.1) and (1.2).
Their form factors will be calculated at maximum momentum
transfer and at leading order in the inverse of the heavy quark mass by
including the contributions of the low lying pole contributions.

\resection{The heavy-light chiral lagrangian}

To be self-contained and to establish the notations
we shall start by reviewing the description
of heavy mesons and light pseudoscalars by effective field operators
and of their effective chiral lagrangian.
Negative parity heavy $Q{\bar q}_a$ mesons are represented by fields
described by a $4 \times 4$ Dirac matrix
\bea
H_a &=& \frac{(1+\slash v)}{2}[P_{a\mu}^*\gamma^\mu-P_a\gamma_5]\\
{\bar H}_a &=& \gamma_0 H_a^\dagger\gamma_0
\eea
Here $v$ is the heavy meson velocity, $a=1,2,3$
(for $u,d$ and $s$ respectively),
$P^{*\mu}_a$ and $P_a$ are annihilation operators normalized as follows
\bea
\langle 0|P_a| Q{\bar q}_a (0^-)\rangle & =&\sqrt{M_H}\\
\langle 0|P^*_a| Q{\bar q}_a (1^-)\rangle & = & \epsilon^{\mu}\sqrt{M_H}
\eea
with $v^\mu P^*_{a\mu}=0$ and $M_H=M_P=M_{P^*}$, the supposedly degenerate
 meson masses.
Also $\slash v H=-H \slash v =H$, ${\bar H} \slash v=
-\slash v {\bar H}={\bar H}$.
The pseudoscalar light mesons are described by
\be
\xi=\exp{\frac{iM}{f_{\pi}}}
\ee
where
\be
{M}=
\left (\begin{array}{ccc}
\sqrt{\frac{1}{2}}\pi^0+\sqrt{\frac{1}{6}}\eta & \pi^+ & K^+\nn\\
\pi^- & -\sqrt{\frac{1}{2}}\pi^0+\sqrt{\frac{1}{6}}\eta & K^0\\
K^- & {\bar K}^0 &-\sqrt{\frac{2}{3}}\eta
\end{array}\right )
\ee
and $f_{\pi}=132 MeV$. Under the chiral symmetry the fields transform as
follows
\bea
\xi & \to  & g_L\xi U^\dagger=U\xi g_R^\dagger\\
\Sigma & \to  & g_L\Sigma {g_R}^\dagger\\
H & \to  & H U^\dagger\\
{\bar H} & \to & U {\bar H}
\eea
where  $\Sigma=\xi^2$, $g_L$, $g_R$ are global $SU(3)$
transformations and $U$ is a
function of $x$, of the fields and of $g_L$, $g_R$.

The lagrangian describing the fields $H$ and $\xi$ and their interactions,
under the hypothesis of chiral and spin-flavour symmetry and at the lowest
order in light mesons derivatives is
\be
\LL_{0}=\frac{f_{\pi}^2}{8}<\partial^\mu\Sigma\partial_\mu
\Sigma^\dagger > +i < H_b v^\mu D_{\mu ba} {\bar H}_a >
+i g <H_b \gamma_\mu \gamma_5 \AA^\mu_{ba} {\bar H}_a>
\ee
where $<\ldots >$ means the trace, and
\bea
D_{\mu ba}&=&\delta_{ba}\partial_\mu+\VV_{\mu ba}
=\delta_{ba}\partial_\mu+\frac{1}{2}\left(\xi^\dagger\partial_\mu \xi
+\xi\partial_\mu \xi^\dagger\right)_{ba}\\
\AA_{\mu ba}&=&\frac{1}{2}\left(\xi^\dagger\partial_\mu \xi-\xi
\partial_\mu \xi^\dagger\right)_{ba}
\eea
Besides chiral symmetry, which is obvious, since, under chiral
transformations,
\bea
D_\mu {\bar H} \to U D_\mu {\bar H} \nn\\
\AA_\mu \to U \AA_\mu U^\dagger
\eea
the lagrangian (2.11) possesses the heavy quark spin symmetry $SU(2)_v$, which
acts as
\bea
H_a &\to& {\hat S} H_a \\
{\bar H}_a &\to& {\bar H}_a {\hat S}^\dagger
\eea
with ${\hat S}{\hat S}^\dagger=1$ and $[\slash v,{\hat S}]=0$, and a heavy
quark flavour symmetry arising from the absence of terms containing
$m_Q$.

Explicit symmetry breaking terms can  also be introduced, by
adding to $\LL_0$
the extra piece (at the lowest order in $m_q$ and $1/m_Q$):
\bea
\LL_1 &=&\lambda_0<m_q\Sigma+m_q\Sigma^\dagger>
+\lambda_1<{\bar H}_a H_b(\xi m_q\xi+\xi^\dagger m_q\xi^\dagger)_{ba}>
\nn\\
&+&\lambda_1^\prime <{\bar H}_a H_a(m_q\Sigma+m_q\Sigma^\dagger)_{bb}>
+\frac{\lambda_2}{m_Q}<{\bar H}_a\sigma_{\mu\nu} H_a\sigma^{\mu\nu}>
\eea
The last term in the previous equation induces a mass difference
between the states $P$ and $P^*$ contained in the field $H$,
such that
\be
 M_P=M_H ~~~~~~~~~~ M_{P^*}=M_H+\delta m_H
\ee
The preceding construction can be found for instance in the paper by
Wise [2], and we have used the same notations.

\resection{Introduction of light vector resonances}

The vector meson resonances belonging to the low-lying $SU(3)$ octet can be
introduced by using the hidden gauge symmetry approach \cite {light}
(for a different approach see \cite{Schec}).
The new lagrangian containing these particles,to be added to $\LL_0+
\LL_1$, is as follows \cite {light}:
\bea
\LL_2&=& -\frac{f^2_{\pi}}{2}a <(\VV_\mu-
\rho_\mu)^2>+\frac{1}{2g_V^2}<F_{\mu\nu}(\rho)F^{\mu\nu}(\rho)> \nn\\
&+&i\beta <H_bv^\mu\left(\VV_\mu-\rho_\mu\right)_{ba}{\bar H}_a>\nn\\
&+&\frac{\beta^2}{2f^2_{\pi} a}<{\bar H}_b H_a{\bar H}_a H_b>+
i \lambda <H_b \sigma^{\mu\nu} F_{\mu\nu}(\rho)_{ba} {\bar H}_a>
\eea
where $F_{\mu\nu}(\rho)=\partial_\mu\rho_\nu-\partial_\nu\rho_\mu+
[\rho_\mu,\rho_\nu]$, and $\rho_\mu$ is defined as
\be
\rho_\mu=i\frac{g_V}{\sqrt{2}}\hat\rho_\mu
\ee
$\hat\rho$ is a hermitian $3\times 3$ matrix analogous to (2.6) containing
the light vector mesons $\rho^{0,\pm}$, $K^*$, $\omega_8$. $g_V$, $\beta$
and $a$ are coupling constants; by imposing the two KSRF relations
\cite {light} one obtains
\be
a=2 \ \ \ \ \ \ \ \ \ \ \ \ \ \ \ g_V \approx 5.8
\ee
We note that the quartic term in the heavy fields $H$ in (3.1) is added to
obtain the simple lagrangian $\LL_0$ in the formal limit $m_{\rho} \to
\infty$, when the $\rho$ field decouples.

\resection{Inclusion of positive parity heavy mesons}

For our subsequent analysis of the heavy mesons
semileptonic decays we shall have to
introduce the low-lying positive parity $Q{\bar q}_a$ heavy meson states.
For $p$ waves ($l=1$), the heavy quark effective theory predicts two
distinct multiplets, one containing a $0^+$ and a $1^+$ degenerate states,
the other one comprising a $1^+$ and a $2^+$ state \cite {IW}, \cite {Ros}.
In matrix notation they are described respectively by \cite {Falk}
\be
S_a=\frac{1+\slash v}{2} \left[D_1^\mu\gamma_\mu\gamma_5-D_0\right]
\ee
and
\be
T_a^\mu=\frac{1+\slash v}{2}\left[D_2^{\mu\nu}\gamma_\nu-\sqrt{\frac{3}{2}}
{\tilde D}_{1 \nu}\gamma_5\left[g^{\mu\nu}-\frac{1}{3}\gamma^\nu (
\gamma^\mu-v^\mu)\right]\right]
\ee
Note that $\slash v S=S \slash v=S$; $\slash v T^\mu=-T^\mu \slash v=
T^\mu$; $\slash v {\bar S} ={\bar S}  \slash v={\bar S}$; $-\slash v
{\bar T}^\mu ={\bar T}^\mu \slash v= {\bar T}^\mu$.
The two multiplets have $s_l=1/2$ and $s_l=3/2$  respectively, where
$\vec {s_l}$ is the angular momentum of the light degrees of freedom
which is conserved together with the heavy quark spin $\vec {s_Q}$
in the infinite quark mass limit because $\vec {J}=\vec{s_l}+\vec{s_Q}$.
The lagrangian containing the fields $S_a$ and $T^\mu_a$ as well as their
interactions with the Goldstone bosons and the fields $H_a$ has been
derived in ref. \cite {Falk}:
\be
\LL_3=\LL_{kin}+\LL_{1\pi}+\LL_s+\LL_d
\ee
\bea
\LL_{kin} & = & i<S_b (v \cdot D)_{ba} {\bar S}_a>+ i<T^\mu_b
(v \cdot D)_{ba} {\bar T}_{\mu a}> \nn\\
&-&\delta m_S <S_a {\bar S}_a>-\delta m_T <T^\mu_a {\bar T}_{\mu a}>
\eea
\be
\LL_{1\pi}=ig'<S_b \gamma_\mu\gamma_5 \AA_{ba}^\mu {\bar S}_a>+
ig''<T^\mu_b \gamma_\lambda\gamma_5 \AA_{ba}^\lambda {\bar T}_{\mu a}>
\ee
\be
\LL_s=i f'<T_b^\mu \AA_{\mu ba}\gamma_5 {\bar S}_a>+i f''<S_b \gamma_\mu
\gamma_5 \AA_{ba}^\mu {\bar H}_a>+ h.c.
\ee
\bea
\LL_d &= & i \frac{h_1}{\Lambda_\chi}<T^\mu_b \gamma_\lambda \gamma_5 (
D_\mu \AA^{\lambda})_{ba} {\bar H}_a>\nn\\
&+& i \frac{h_2}{\Lambda_\chi}<T^\mu_b \gamma_\lambda \gamma_5 (
D^\lambda \AA_{\mu})_{ba} {\bar H}_a>
\eea
In (4.4) $\delta m_S=M_{D_0}-M_H=M_{D_1}-M_{H}$,
$\delta m_T=M_{D_2}-M_H=M_{{\tilde D}_1}-M_{H}$ .
Notice that a mixing term between the $S$ and $T_\mu$ fields
is absent at the leading order. Indeed, by saturating the
$\mu$ index of $T_\mu$ with $v_\mu$ or $\gamma_\mu$ gives a
vanishing result, and derivative terms are forbidden by the
reparametrization invariance \cite{Falk}.

We add here the coupling of the vector meson light resonances to the
positive and negative parity states
\be
\LL_4=\LL_{S\rho}+\LL_{T\rho}+\LL'
\ee
\be
\LL_{S\rho} = i \beta_1<S_b v^\mu (\VV_\mu-\rho_\mu)_{ba} {\bar S}_a>
+i \lambda_1 <S_b \sigma^{\mu\nu}F_{\mu\nu}(\rho)_{ba}{\bar S}_a>
\ee
\be
\LL_{T\rho} = i \beta_2 <T^{\lambda}_b v^\mu (\VV_\mu -\rho_\mu )_{ba}
{\bar T}_{a \lambda}> +i \lambda_2 <T^{\lambda}_b \sigma^{\mu\nu}
F_{\mu\nu}(\rho)_{ba}{\bar T}_{a \lambda}>
\ee
\bea
\LL' & = & i \zeta < {\bar S}_a H_b \gamma_\mu (\VV^\mu-\rho^\mu)_{ba}>
+i \mu <{\bar S}_a H_b\sigma^{\lambda\nu} F_{\lambda\nu}(\rho )_{ba}
> \nn \\
&+&i \zeta_1 < {\bar H}_a T^{\mu}_b \gamma_\mu
(\VV^\mu -\rho^\mu)_{ba}> +i \mu_1 <{\bar H}_a T^{\mu}_b \gamma^\nu
 F_{\mu\nu}(\rho )_{ba}>
\eea
We shall see in the following that some information on the coupling
constants $g$, $\mu$, $\lambda$ and $\zeta$ can be obtained by the analysis
of the semileptonic decays (1.1) and (1.2).

\resection{Weak currents}

At the lowest order in derivatives of the pseudoscalar couplings and in the
symmetry limit, weak interactions between light pseudoscalars and a
heavy meson are described
by the weak current \cite {Wise}:
\be
L^{\mu}_a=\frac{i\alpha}{2} <\gamma^{\mu} (1-\gamma_5) H_b \xi^{\dagger}_{ba}>
\ee
where $\alpha$ is related to the pseudoscalar heavy meson decay constant $f_H$,
defined by
\be
<0|\overline{q}_a \gamma^{\mu} \gamma_5 Q|P_b (p)>=ip^{\mu} f_H \delta_{ab}
\ee
as follows:
\be
\alpha=f_H \sqrt{M_H}
\ee
We can in a similar way introduce the current describing the weak interactions
between pseudoscalar Goldstone bosons and the positive parity $S$ fields:
\be
\hat {L}^{\mu}_a = \frac {i\hat{\alpha}} {2} <\gamma^{\mu} (1-\gamma_5) S_b
\xi^{\dagger}_{ba}>
\ee
and the current by which the H fields interact with the light vector mesons:
\be
L_{1 a}^{\mu}=\alpha_1 <\gamma_5 H_b (\rho^{\mu}-V^{\mu})_{bc}
\xi^{\dagger}_{ca}>
\ee
All these currents transform under the chiral group similarly to the quark
current $\overline{q} \gamma^{\mu} (1-\gamma_5)Q$, i.e. as $(\overline{3}_L,
1_R)$. We also observe that there is no similar coupling between the fields
$T^{\mu}$ and $\xi$. Indeed (5.1) and (5.4) also describe the matrix element
between the meson and the vacuum, and this coupling vanishes for the $1^+$ and
$2^+$ states having $s_{l}=3/2$. This can be proved explicitly by
considering the current matrix element ($A^{\mu}=\overline{q}_a \gamma^{\mu}
\gamma_5 Q$):
\be
<0|A^{\mu}|\tilde{D}_1> = \tilde{f} \epsilon^{\mu}
\ee
Using the heavy quark spin symmetry and the methods of
the first two papers in ref. \cite {Isgur},
(5.6) turns out to be proportional to the matrix element of the vector
current between the vacuum and the $2^+$ state, which vanishes.

\resection{Semileptonic decays}

Let us first consider the decay (1.1). The hadronic matrix element
can be written in terms of the form factors $F_0$, $F_1$ as follows
\be
<\Pi(p')|V^{\mu}|P(p)> =
 \big[ (p+p')^{\mu}+\frac{M_\Pi^2-M_H^2}{q^2} q^{\mu}\big]
F_1(q^2) -\frac {M_\Pi^2-M_H^2}{q^2} q^{\mu} F_0(q^2)
\ee
where $q^{\mu}=(p-p')^{\mu}$, $F_0(0)=F_1(0)$ and $M_H=M_P$ (see (2.18)).
The form factors $F_0$ and $F_1$
take contributions, in a dispersion relation, from the $0^+$ and $1^-$ meson
states respectively.

We notice here that, by working at the leading order in $1/m_Q$, the
possible parametrizations of the weak current matrix element are not
all equivalent. Computed in the heavy meson effective theory, the
matrix element of eq. (6.1) reads:
\be
<\Pi(p')|V^{\mu}|P(p)> =
Av^\mu+ B {p^\prime}^\mu
\ee
with $A$ and $B$ both scaling as $\sqrt{M_H}$ at $q^2=q^2_{max}=(M_H-M_\Pi)^2$
(where the theory should provide for a better approximation).
The factor $\sqrt{M_H}$ which gives rise to this scaling behaviour comes
just from the wave function normalization of the $P$ operator, and no other
explicit factor $M_H$ appears in the heavy meson effective field theory.
If one
introduces the usual form factors $f_+$ and $f_-$ through the following
decomposition:
\be
<\Pi(p')|V^{\mu}|P(p)> =
f_+ (p+p^\prime)^\mu+f_-(p-p^\prime)^\mu
\ee
one has the relations:
\be
f_+=\frac{1}{2}\left(\frac{A}{M_H}+B\right),~~~~~
f_-=\frac{1}{2}\left(\frac{A}{M_H}-B\right)
\ee
It would seem consistent at this point to throw away the terms proportional
to $A$, obtaining
\be
<\Pi(p')|V^{\mu}|P(p)>
\simeq B{p^\prime}^\mu
\ee
which however does not reproduce the original expression of the matrix
element. This is a clear contradiction since the two terms on the right hand
side of eq. (6.2) scale in the same fashion. On the other hand, by making
use of the decomposition of eq. (6.1) and working at the leading order
we find:
\be
F_1=\frac{B}{2},~~~~~~~~~~F_0=\frac{1}{M_H}\left(A+B M_\Pi\right)
\ee
which, inserted back in the eq. (6.1), fully reproduces the matrix
element given in eq. (6.2). The previous example shows that one must be
very careful in the definition of the form factors when working at
the leading order in $1/m_Q$ in the heavy meson effective field theory.

Using the previous lagrangians (2.11), (4.6) and the currents (5.1), (5.4)
 we obtain,
at the leading order in $1/m_Q$ and at $q^2=q^2_{\rm max}$,
the following results
\be
F_1(q^2_{\rm max})=\frac {g M_H f_H} {2 f_{\pi} (v\cdot k -\delta m_H)}
\ee
\be
F_0(q^2_{\rm max})=\frac {f'' \hat {\alpha} M_\Pi } { \sqrt{M_H} f_{\pi}
(v\cdot k -\delta m_S)} -\frac{f_H}{f_{\pi}}
\ee
The r.h.s. in (6.7) and the first term in (6.8) arise from polar diagrams.
Finally $k^{\mu}$ is the residual momentum
related to the physical momenta by $k^{\mu}=q^{\mu}-M_{\tilde H} v^{\mu}$ (and
$p^{\mu}=M_H v^{\mu}$).

A similar analysis can be performed for the semileptonic decay process (1.2)
of a heavy pseudoscalar meson $P$
with a light vector $\Pi ^*$ particle in the final state.
The current matrix element
is expressed as follows
\bea
<\Pi^* (\epsilon,p')| &(&V^{\mu}-A^{\mu})|P(p)> =
\frac {2 V(q^2)} {M_H+M_{\Pi^*}}
\epsilon^{\mu \nu \alpha \beta}\epsilon^*_{\nu} p_{\alpha} p'_{\beta} \nn\\
&+& i  (M_H+M_{\Pi^*})\left[ \epsilon^*_\mu -\frac{\epsilon^* \cdot q}{q^2}
q_\mu \right] A_1 (q^2) \nn\\
& - & i \frac{\epsilon^* \cdot q}{(M_H+M_{\Pi^*})} \left[ (p+p')_\mu -
\frac{M_{H}^2-M_{\Pi^*}^2}{q^2} q_\mu \right] A_2 (q^2) \nn \\
& + & i \epsilon^* \cdot q \frac{2 M_{\Pi^*}}{q^2} q_\mu A_0 (q^2)
\eea
where
\be
A_0 (0)=\frac {M_{\Pi^*}-M_H} {2M_{\Pi^*}} A_2(0) + \frac {M_{\Pi^*}+M_H}
{2M_{\Pi^*}} A_1(0)
\ee
Notice that the tensor structures given in square brackets of eq. (6.9)
have vanishing divergence and are constant in the limit of infinite
$M_H$.
Such a decomposition satisfies the same properties discussed above for
the form factors $F_0$ and $F_1$.
In a dispersion relation the form factor $V(q^2)$ takes contribution from $1^-$
particles, $A_0(q^2)$ from $0^-$ particles and $A_j(q^2)$ ($j=1,2$) from $1^+$
states.

Using the lagrangians (3.1) and (4.11) and the currents (5.1), (5.4) and
(5.5) we get
at $q^2=q^2_{\rm max}$ and at leading order in $1/m_Q$ the results
\be
V(q^2_{\rm max})=-\frac {g_V} {\sqrt 2} \lambda f_H
\frac {M_H+M_{\Pi^*}} {v\cdot
k -\delta m_H}
\ee
\be
A_1(q^2_{\rm max})= - {\frac {2 g_V}{\sqrt 2}} \left [ {\frac
{\alpha_1 \sqrt{M_H}}{M_H +M_{\Pi^*}}} + {\frac {\hat {\alpha} \sqrt{M_H}}
{M_H +M_{\Pi^*}}} {\frac {\zeta/2 -\mu M_{\Pi^*}}{v\cdot k -\delta m_S}}
\right ]
\ee
\be
A_2(q^2_{\rm max})={\frac {\mu g_V}{\sqrt 2}}
 \frac{{\hat \alpha}}{\sqrt{M_H}} {\frac {M_H+M_{\Pi^*}}
{v\cdot k -\delta m_S}}
\ee
\be
A_0(q^2_{\rm max})=-{\frac {g_V}{2 \sqrt 2}} {\frac {\beta f_H M_H}
{M_{\Pi^*} (v\cdot k -\delta m')}} + \frac {g_V} {\sqrt 2} \frac {\sqrt {M_H}}
{M_{\Pi^*}} \alpha_1
\ee
where $\delta m'$ arise from the chiral breaking terms of Eq.(2.17).
The first term in (6.12)  and the last one in (6.14)
arises from the direct coupling between the heavy
meson $H$ and the $1^-$ light resonances of Eq.(5.5) and the other ones
{}from polar diagrams.

\resection{Numerical analysis}

The results (6.7),(6.8) and (6.11)-(6.14) are obtained in the chiral limit
and for
$m_Q \to \infty$; therefore they should apply (with non-leading
corrections) to the
decays $B \to \pi \ell \nu_{\ell}$ or $B \to \rho \ell \nu_{\ell}$.
Unfortunately, for those decays there are not sufficient
 experimental results that could be
used to determine the various coupling constants appearing in the final
formulae.

On the other hand, for $D$ decays the experimental information is much more
detailed and we could tentatively try to use it to fix the constants as well as
to make predictions on the other decays which have not been measured yet.

In order to make  contact with the experimental data, we have to know the
behaviour of the form factors with $q^2$. Except for the direct terms in (6.8),
 (6.12) and (6.14)
 all the contributions we have collected arise from polar diagrams,
which suggests a simple pole behaviour. This is also the assumption usually
made in the phenomenological analysis of $D$ semileptonic decays.
Therefore we have
assumed for the form factors $F_1(q^2)$, $V(q^2)$, $A_1(q^2)$ and $A_2(q^2)$
(the form factors $F_0(q^2)$ and $A_0(q^2)$ are not easily accessible to
measurement since they appear in the width multiplied by the lepton mass) the
generic formula
\be
F(q^2)= \frac {F(0)} {1 - \frac {q^2}{m^2}}
\ee
For the pole masses we use the inputs in Table I \cite {Korner} that also
agree with the masses fitted by the experimental analyses of $D$ decays
\cite {Stone}.

For the $D \to \pi$ semileptonic decay one thus gets,
{}from (6.7) and (7.1):
\bea
F_1(0)&=& -\frac{g \alpha}{2 f_{\pi}} \sqrt{M_D} \;\frac {M_{D^*}+M_D -M_{\pi}}
{M_{D^*}^2}\\
&=& -\frac{g f_D }{2 f_{\pi}} \frac {M_D (M_{D^*}+M_D -M_{\pi})}{M_{D^*}^2}
\eea
For $f_D$ we use the value suggested by lattice \cite {Gavela} QCD and by
QCD sum rules analysis \cite {[14]}, $f_D=200 MeV$.

Experimentally one has \cite {pdb} $|F_1(0)|=0.79\pm 0.20$, which implies
\be
|g|=0.61\pm 0.22
\ee
This result agrees with the result that would have been obtained using as input
$D\to K$ semileptonic decay \cite{Stone}:
 $|g|=0.57\pm0.13$ and is also in
agreement with the result obtained by a recent analysis of radiative $D^*$
decays: $|g|=0.58\pm 0.41$ (for $m_c=1700 MeV$) \cite {Georgi}.

Let us now turn to semileptonic decays into vector mesons. The experimental
inputs we can use are from $D \to K^* \ell \nu_{\ell}$ and are as follows:
\bea
V(0)&=&0.95\pm 0.20\nn\\
A_1(0)&=&0.48\pm 0.05\nn\\
A_2(0)&=&0.27\pm 0.11
\eea
They are  averages between the data from E653 \cite {E653} and E691 \cite
{E691} experiments. The calculated weak couplings at $q^2=0$ are:
\bea
V(0)& = & \frac {g_V \lambda}{\sqrt 2} \frac {(M_D+M_{K^*}) (M_{D^*}
+M_D-M_{K^*})}{M_{D^*}^2} \frac {\alpha}{\sqrt{M_D}}
\nn \\
& = & \frac {g_V \lambda}{\sqrt 2} \frac {(M_D+M_{K^*})
(M_{D^*}+M_D-M_{K^*})}{M_{D^*}^2} f_D
\eea
\bea
A_1(0)&=&-\sqrt{2} g_V \frac {(M_{D_1}+M_D-M_{K^*}) \sqrt{M_D}} {(M_D+M_{K^*})
M_{D_1}^2} \times\nn\\
& &\left [ \alpha_1 (M_{D_1}-M_D+M_{K^*})-\hat{\alpha} (\frac
{\zeta}{2} -\mu M_{K^*}) \right ]
\eea
\be
A_2(0)=-\frac {g_V \mu}{\sqrt 2} \frac {(M_D+M_{K^*}) (M_{D_1}+M_D-M_{K^*})}
{M_{D_1}^2} \frac {\hat \alpha}{\sqrt{M_D}}
\ee
Taking $f_D=200 MeV$, from Eq.(7.5), (7.6) and (7.8) we obtain:
\bea
|\lambda|&=&0.60\pm 0.11 ~GeV^{-1}\\
\hat {\alpha}\; \mu &=& -0.06\pm 0.02 ~GeV^{3/2}
\eea
By using the result $\hat {\alpha} = 0.46\pm 0.06 ~GeV^{3/2}$ from QCD sum
rules \cite {Beppe}, one obtains:
\be
\mu=-0.13\pm 0.05
\ee
For the $A_1$ coupling the experimental data do not allow a separate
determination of $\alpha_1$ and $\zeta$. However we notice that the
combination:
\be
\alpha_{eff} = \alpha_1 (M_{D_1}-M_D+M_{K^*}) -{\hat \alpha} \left ( \frac
{\zeta}{2} - \mu M_{K^*}\right )
\ee
is almost flavour independent and, at leading order in the $1/M_Q$ expansion
is scaling invariant. From the $D\to K^*$ data given in Eq.(7.5) we find:
\be
\alpha_{eff} = -0.22\pm 0.02 ~GeV^{3/2}
\ee
We can now give predictions for the processes which the heavy quark and chiral
symmetries relate to $D \to \pi$ and $D \to K^*$. Concerning $D \to \pi$, we
can use Eq.(7.3) together with the value of the constant $g$ given in
Eq.(7.4) to derive $F_1(0)$ for the various decays. Taking $f_B=f_D=200
MeV$ we obtain the results given in Table II.
Notice that, by using $f_B=f_D$ in Eq.(7.3), we are
implicitly accounting for the large corrections to the relation
$f_B/f_D=\sqrt{M_D}/\sqrt{M_B}$, which is implied by  lattice QCD and QCD
sum rules results\footnote{After completion of this work
we received a paper by Burdman \cite{Burdnew}. There, a formal argument is
provided which supports the idea that in the semileptonic transition of the
kind $P \to \Pi$ the non-leading corrections are mainly reabsorbable
in the $f_D$ decay constant. A straightforward extension of the
argument to the transition of the kind $P \to \Pi^*$ seems problematic.}
 \cite{Beppe}.
 Had we insisted in using the leading order expression of
our computation, Eq.(7.2), we would have obtained the results shown in
parenthesis in Table II, by fixing from the $D \to \pi$ data the product
$g \alpha$. These results agree with the previous ones in the
$D$ sector, but they obviously disagree for the $B$, predicting partial widths
which are smaller by almost a factor of 3.

For the decays which are related to $D \to K^*$ the situation is more complex.
We have not determined all relevant couplings of the effective lagrangian, from
the $D \to K^*$ data. In particular we have determined a combination of
$\alpha_1$ and $\zeta$, called $\alpha_{eff}$ and given in Eq.(7.12).
In the
expression of $V(0)$ we shall still choose $f_D=f_B=200 MeV$, in agreement
with the lattice and sum rules calculations.

This approach leads to the results given in Table III, expressed as predictions
for the transverse, longitudinal and total widths $\Gamma_T$, $\Gamma_L$ and
$\Gamma$. For comparison we have also displayed in parenthesis the results
obtained by working strictly at the leading order in $1/m_Q$, avoiding the
identifications $\alpha=f_D \sqrt{M_D}$ and
fitting from the $D\to K^*$ data the combinations $\lambda \alpha$,
$\alpha_{eff}$ and $\hat {\alpha} \mu$. The predictions for the form factors
$A_1(0)$ and $A_2(0)$ are in this case the same and, as a consequence, the
predicted values for $\Gamma_L$ coincide for all the considered decays. On the
other hand $V(0)$ for the $B$ decays is smaller if computed at the leading
order. This implies a transverse width $\Gamma_T$ smaller by a factor two and a
total width $\Gamma$ smaller by about a factor 1.6.

The results of Table III cannot be fully compared to experiments due to the
lack of data. For the decay $D^+ \to \rho^0 \ell^+ \nu_{\ell}$ one has
the upper limit \cite{pdb} $BR < 3.7 \cdot 10^{-3}$, which is satisfied by our
result $BR(D^+ \to \rho^0 \ell^+ \nu_{\ell})=2.4 \cdot 10^{-3}$.

For the decay $B^- \to \rho^0 \ell^- {\bar \nu}_{\ell}$ we obtain
$BR= 0.44 \cdot 10^{-3}$ (resp. $0.28 \cdot 10^{-3}$ in the leading order
approximation for $f_B$ and $f_D$), to be compared with the ARGUS result
\cite{argus}: $BR= (1.13 \pm 0.36 \pm 0.26) \cdot 10^{-3}$, which however is
not confirmed by CLEO collaboration \cite{cleo} that finds an upper limit of
about $0.3 \cdot 10^{-3}$.

It is curious to observe that the leading order results could have been
obtained in a model independent way by assigning, in the parametrization of the
matrix element, the scaling behaviour of the various form factors. For
instance, for the $D \to K^*$ process we can write:
\bea
\frac {V} {(M_D+M_{K^*})} &=& \frac {v} {\sqrt {M_D}}\\
(M_D+M_{K^*}) A_1 &=& a_1 \sqrt{M_D}\\
\frac {A_2} {(M_D+M_{K^*})} &=& \frac {a_2}{\sqrt{M_D}}
\eea
where $v$, $a_1$ and $a_2$ are constants as $M_D$ grows. This behaviour simply
follows from the definitions of $V$, $A_1$ and $A_2$, and from the fact that
the matrix element $<K^*|J^{\mu}|D>$ scales as $\sqrt{M_D}$. The above
relations are valid at $q^2=q^2_{\rm max}=(M_D-M_{K^*})^2$ and they should be
appropriately modified at $q^2=0$. To do so we assume a simple polar behaviour
for the form factors. Notice that the quantities $v$, $a_1$ and $a_2$ will in
general depend on $M_D$, $M_{K^*}$ and the relevant pole mass $M_{\rm Pole}$,
with the restriction that they should be constant in the large $M_D$ limit. At
$q^2_{\rm max}$ the polar behaviour provides a factor:
\be
\frac {M^2_{\rm Pole}}{M^2_{\rm Pole}-(M_D-M_{K^*})} \sim \frac {1}{2}
M_{\rm Pole} \frac {1} {(M_{\rm Pole}-M_D+M_{K^*})}
\ee
This factor exhibits a certain flavour dependence, which we may account for by
incorporating it in $v$, $a_1$ and $a_2$:
\be
v=\frac {\hat {v}} {(M_{\rm Pole}-M_D+M_{K^*})}
\ee
and similarly for $a_1$, $a_2$. We can assume that $\hat v$, $\hat {a_1}$ and
$\hat {a_2}$ are approximately flavour independent.
In this way we obtain the following expressions
\bea
V(0)&=& {\frac {(M_D+M_{K^*}) (M_{\rm Pole}+M_D-M_{K^*})} {M^2_{\rm Pole}
\sqrt {M_D}}} \hat {v}\\
A_1(0)&=& {\frac {(M_{\rm Pole}+M_D-M_{K^*}) \sqrt{M_D}} {(M_D+M_{K^*})
M^2_{\rm Pole}}} \hat {a_1}\\
A_2(0)&=& {\frac {(M_D+M_{K^*}) (M_{\rm Pole}+M_D-M_{K^*})} {M^2_{\rm Pole}
\sqrt {M_D}}} \hat {a_2}
\eea
The constants $\hat v$, $\hat {a_1}$ and $\hat {a_2}$ are determined by the
data for $D \to K^*$ given in Eq.(7.5).

A comparison with our model gives:
\bea
{\hat v} &=& \frac {g_V \lambda}{\sqrt 2} \alpha\\
\hat {a_1} &=& -\sqrt{2} g_V \alpha_{eff}\\
\hat {a_2} &=& -\frac{g_V \mu}{\sqrt 2} \hat{\alpha}
\eea
therefore the predictions obtained from this scaling argument coincide with
those obtained at leading order from an effective lagrangian.

In  Table IV we compare our results, for $f_D=f_B=200 MeV$ with
other existing calculations. The comparison is made for the ratios of the form
factors at $q^2=0$ to the corresponding form factors for the $D$ meson, from
which we have fixed our parameters.

\resection{Conclusions}

The leptonic decays of a heavy pseudoscalar meson into a light pseudoscalar or
into an octet vector resonance have been studied with our effective lagrangian
by including the allowed direct coupling and the lowest contributing poles. The
formalism can be reliable only at $q^{2}_{max}$ and to leading order in
$1/m_Q$. Most of the experimental information is available only for $D$ decays.
To extract information at other momentum transfers one has to assume generic
pole extrapolations. In this we follow the experimental phenomenological
analyses. From the present data on semileptonic decay of $D$ into pion, and by
using $f_D =200~MeV$, we can extract for the coupling constant $g$ appearing in
the effective chiral coupling of pseudoscalars with heavy mesons a value
$|g|=0.61 \pm 0.22$ in agreement with those obtained from radiative $D^*$
decays (and also from decay into $K$). We can then try to predict the branching
ratios for the related decays $D \to K$, $D \to \eta$, $D_s \to \eta$, $D_s \to
K$, $B \to \pi$, $B_s \to K$, as shown in Table II. A similar analysis for the
decays related to $D \to K^*$ through heavy quark and chiral symmetries
requires additional assumptions to arrive at the predictions shown in Table III
for the transverse, longitudinal, and total widths. For the $D$ decays in that
table one can develop a scaling argument leading essentially to the same
predictions. On the other hand the numerical estimates for $B \to$ vector
resonance with the dynamical model based on the effective lagrangian differ
considerably from those of such a scaling argument, as also shown in Table III.
Our predictions can be compared with those of other calculations in the
literature. The comparison can be made in terms of the ratios of the form
factors at vanishing momentum transfer to the $D$ meson corresponding form
factors used to fix the parameters. Significant differences are noticed
among different models, which
shows the still uncertain status of
the theory. The theoretical analysis we have presented here is based on a
dynamically structured approach, using an effective lagrangian including the
presumably relevant degrees of freedom. The existing data would not leave much
space for a more accurate treatment by including non-leading contributions.
Under such a limitation the model allows for predictions for the $D \to$
pseudoscalar leptonic decays, for the $D \to$ light vector resonance and,
probably with some more uncertainties, for the $B \to $ light vector resonance
leptonic decays.

The present status of the subject, in particular the still insufficient
experimental data, do not yet allow for a more complete theoretical approach of
a precision comparable to that of low energy applications of chiral lagrangians
\cite {Leutwyler}. In this sense the calculations presented here are to
be considered as still exploratory.
Additional experimental data would greatly help
in a better determination of the parameters of the effective lagrangian
proposed here and in testing for the possible necessity of non-leading
corrections.

\newpage
\begin{center}
  \begin{Large}
  \begin{bf}
  Tables Captions
  \end{bf}
  \end{Large}
\end{center}
  \vspace{5mm}
\begin{description}
\item [Table I] Pole masses for different states. Units are $GeV$.
\item[ Table II ] Predictions for semileptonic $D$ and $B$ decays in
 a pseudoscalar meson. We have neglected the $\eta ~- \eta '$ mixing.
  The branching ratios and
the widths for $B$ must be multiplied for $|V_{ub}/0.0045|^2$. In the first
column $f_D=f_B=200~ MeV$ is assumed. In parenthesis the leading
order result is assumed, i.e. $f_B/f_D= \sqrt{M_D/M_B}$. We also assume
$\tau_{B_s}=\tau_{B^0} =\tau_{B^+}= 1.29~ps.$
\item[ Table III] Predictions for semileptonic $D$ and $B$ decays
into a vector meson. Partial widths  are in units of $10^{11}~
s^{-1}$ . The branching ratios and
the widths for $B$ must be multiplied for $|V_{ub}/0.0045|^2$. The first
column refers to the case $f_B=f_D=200~ MeV$.
The results in parenthesis have been obtained
in the leading order.
\item [Table IV] Comparison  among our predictions
and other theoretical calculations of the form factors at $q^2=0$.
The results in parenthesis have been obtained
in the leading order.
\end{description}
\end{document}